\documentclass[twocolumn]{aastex63}

\usepackage{soul}
\usepackage{multirow}
\usepackage{booktabs}

\hypersetup{linkcolor=red,citecolor=green,filecolor=cyan,urlcolor=magenta}

\newcommand{\pcm}{\,pc cm$^{-3}$}	
\newcommand{\radm}{\,rad m$^{-2}$} 

\received{XXX}
\revised{XXX}
\accepted{XXX}
\submitjournal{ApJ}

\shorttitle{Individual pulse emission from PSR J1741$-$0840}
\shortauthors{Y. H. Xu et al.}

\begin{document}

\title{Investigation of individual pulse emission behaviours from pulsar J1741$-$0840}

\correspondingauthor{Z. G. Wen}
\email{wenzhigang@xao.ac.cn}

\author{Y. H. Xu}
\affiliation{Yunnan Observatories, Chinese Academy of Sciences,
Kunming, Yunnan, 650011, People's Republic of China}

\author{Z. G. Wen}
\affiliation{Xinjiang Astronomical Observatory, Chinese Academy of Sciences,
Urumqi, Xinjiang, 830011, People's Republic of China}

\author{J. P. Yuan}
\affiliation{Xinjiang Astronomical Observatory, Chinese Academy of Sciences,
Urumqi, Xinjiang, 830011, People's Republic of China}

\author{Z. Wang}
\affiliation{Xinjiang Astronomical Observatory, Chinese Academy of Sciences,
Urumqi, Xinjiang, 830011, People's Republic of China}
\affiliation{Xinjiang Key Laboratory of Radio Astrophysics, Urumqi, Xinjiang,
830011, People's Republic of China}

\author{X. F. Duan}
\affiliation{Xinjiang Astronomical Observatory, Chinese Academy of Sciences,
Urumqi, Xinjiang, 830011, People's Republic of China}
\affiliation{Key Laboratory of Microwave Technology, Urumqi, Xinjiang, 830011,
People's Republic of China} 

\author{Z. Wang}
\affiliation{Xinjiang Astronomical Observatory, Chinese Academy of Sciences,
Urumqi, Xinjiang, 830011, People's Republic of China}
\affiliation{School of Astronomy and Space Sciences, University of Chinese Academy of Sciences, Beijing, 100049, People's Republic of China}

\author{N. Wang}
\affiliation{Xinjiang Astronomical Observatory, Chinese Academy of Sciences,
Urumqi, Xinjiang, 830011, People's Republic of China}

\author{M. Wang}
\affiliation{Yunnan Observatories, Chinese Academy of Sciences,
Kunming, Yunnan, 650011, People's Republic of China}

\author{H. G. Wang}
\affiliation{School of Physics and Materials Science, Guangzhou University,
Guangzhou, Guangdong, 510006, People's Republic of China}

\author{A. Rusul}
\affiliation{Institute of Physics and Electrical Engineering, Kashi University,
Kashgar, Xinjiang, 844009, People's Republic of China}

\author{L. F. Hao}
\affiliation{Yunnan Observatories, Chinese Academy of Sciences,
Kunming, Yunnan, 650011, People's Republic of China}

\author{W. Han}
\affiliation{Xinjiang Astronomical Observatory, Chinese Academy of Sciences,
Urumqi, Xinjiang, 830011, People's Republic of China}

\begin{abstract}
We have carried out a detailed study of individual pulse emission from the
pulsar J1741$-$0840 (B1738$-$08), observed using the Parkes and Effelsberg radio
telescopes at the $L$ band.
The pulsar exhibits four emission components which are not well resolved
by employing multi-component Gaussian fitting.
The radio emission originates at a height of approximately 1000 km,
with the viewing geometry characterized by inclination and impact
angles roughly estimated at 81$^\circ$ and 3$^\circ$, respectively.
Fluctuation spectral analysis of single pulse behaviour reveals two
prominent periodicities, around 32 and 5 rotation periods.
The longer periodic modulation feature is linked to nulling behaviour
across the entire emission window, with an updated nulling fraction
of 23$\pm$2\% is derived from pulse energy distribution via Gaussian
mixture modeling.
In addition to quasiperiodic nulling, the pulsar also exhibits the
presence of subpulse drifting in the trailing component, with the shorter
periodic feature in the fluctuation spectra related to the phenomenon of
subpulse drifting, and the longitudinal separation estimated to be about 5
degrees.
Both periodic modulations show significant temporal evolution with
time-dependent fluctuation power.
The ramifications for understanding the radio emission mechanisms are discussed.
\end{abstract}


\section{Introduction}
\label{sec:intro}
In addition to rotational periodicity, the radio emission from some pulsars
presents periodic modulations in their single-pulse sequences over timescales
ranging from a few seconds to several minutes.
The subpulses appear at progressively shifting longitudes within the pulse
window in successive pulses and the phenomenon is referred to as subpulse
drifting \citep{Drake+Craft+1968}.
The marching of the subpulses can occur in either direction across the emission
window.
Two periodicities are generally adopted to characterize the subpulse drifting:
$P_2$ (the longitudinal spacing between subpulses within a single pulse) and
$P_3$ (the interval over which the drift brings successive subpulses to the same
longitude).
Thus, the drift rate is derived to be $D=P_2/P_3$.
\citet{Weltevrede+etal+2006,Weltevrede+etal+2007} studied the drifting subpulses
for a large sample of unbiased pulsars at both 1428 MHz (21 cm) and 326 MHz
(92 cm) using the Westerbork Synthesis Radio Telescope (WSRT) and concluded
that subpulse drifting is a common behaviour in pulsars.
It is estimated that at least half of the total population of pulsars possess
drifting subpulses when observed with sufficiently high signal-to-noise ratio.
The phenomenon of pulse nulling is observed as the absence of radio emission
temporarily for one or more rotations \citep{Backer+1970} and is highly
concurrent across wide frequency range \citep{Gajjar+etal+2014,Chen+etal+2023}.
The broadband nature implies that the nulling invokes changes on the global
magnetospheric scale.
Recently, \citet{Konar+Deka+2019} compiled the first comprehensive list of nulling
pulsars, revealing that approximately 8 per cent of the pulsar population
exhibit pulse nulling phenomenon, and that these pulsars are statistically
different from regular pulsars.
The physical origins of nulling remain mysterious since its discovery and has
been suspected to be indicative of an aging or dying pulsar.
To quantify the nulling behaviour, the nulling fraction (NF) is used to describe
the percentage of time when there is no discernible emission.
The NF are found to range from less than 1\% up to 95\% or so, and a weak
positive correlation between characteristic age and NF is found
\citep{Ritchings+1976,Wang+etal+2007}.
However, \citet{Konar+Deka+2019} and \citet{Sheikh+MacDonald+2021} found no
correlation in the current data and argued that any apparent non-linear
correlations could be attributed to selection effects.

The subpulse drifting and pulse nulling remain poorly understood both observationally
and theoretically even though they are suggested to be intrinsic properties of
the emission mechanism.
A link between subpulse drifting and nulling behaviour is suggested up the
discovery of quasiperiodic nulling in pulsar B1133+16
\citep{Herfindal+Rankin+2007}.
The occurrence of nulls is observed to be periodic in nature usually with short
durations, which may be governed by emission cycles \citep{Herfindal+Rankin+2009}.
The subpulse drifting is confined to the behaviour of cone emission, while the
periodic nulling manifests across the entire emission window.
Additionally, both periodic modulation features show different distributions of
spin-down energy loss ($\dot{E}$).
The drifting periodicity is weakly anticorrelated with $\dot{E}$
\citep{Basu+etal+2016}, while the periodic nulling does not show any dependence
on $\dot{E}$ \citep{Basu+etal+2017}.
These distinctions between subpulse drifting and periodic nulling signifies their
different emission mechanisms.
These statements are specific to certain characteristics of nulling with
varying durations and therefore cannot be generalized to all nulling and
subpulse drifting phenomena.

In certain pulsars, the phenomena of subpulse drifting and quasiperiodic nulling
are co-existed, and the periodicity of subpulse drifting is much shorter than
nulling periodicity \citep{Basu+etal+2020}.
The interactions between subpulse drifting and nulling are investigated.
For instance, the drift mode transitions in PSRs B0809+74 and B0818$-$13 are
detected to be interspersed with the nulls
\citep{Lyne+Ashworth+1983}.
Furthermore, the phase memory effect is found during these interactions.
For instance, the phase information of the least active pulse is retained during
the null state for PSR B0031$-$07 \citep{Joshi+Vivekanand+2000}.
\citet{Grover+etal+2024} found that the majority of these pulsars tend to be in
the death valley in the period and period derivative diagram
\citep{Konar+Deka+2019}.

The study of quasiperiodic nulling and subpulse drifting offer an invaluable
laboratory to study pulsar emission mechanisms and magnetosphere.
Pulsar J1741$-$0840 is an isolated normal pulsar discovered in an extensive
pulsar survey using the Molonglo radio telescope \citep{Manchester+etal+1978}.
It has a rotational period of $P_1=2.04$ s and a first-period derivative of
$\dot{P_1}=2.27\times10^{-15}$ s s$^{-1}$, giving the characteristic age of
$\tau_c=14.20$ Myr and the surface magnetic field of $B_s=2.18\times10^{12}$ G.
The parallax and proper motion measurements using the Very Long Baseline
Array provide model-independent estimates of distance and transverse velocity,
which are 3.58 kpc and 116.7 km/s, respectively \citep{Deller+etal+2019}.
Good quality profiles were obtained in a short integration time at multiple
frequency bands \citep{Wu+etal+1993,Seiradakis+etal+1995,Kijak+etal+1998,Gould+Lyne+1998,Johnston+Kerr+2018}.
The pulsar's profile comprises of four emission components, which can be
interpreted as a sightline traverse cutting across two concentric rings of inner
and outer cones \citep{Basu+etal+2021}.
And the difference in spectral index between the core and the conal components was
estimated to be $0.363\pm0.095$.
The frequency evolution of the profile reveals that the pulse width decreases
with increasing frequency \citep{Gould+Lyne+1998}.
The central core component is not apparent from the observations at 225 MHz and
325 MHz \citep{Krzeszowski+etal+2009}.
The linear polarization position angle presents well-defined S-shaped swing
observed with the MeerKAT telescope from 896 to 1671 MHz, implying that the
underlying mechanism is coherent curvature radiation \citep{Johnston+etal+2024}.
The flux density follows a power-law spectrum with a spectral index of
$-1.66\pm0.51$ \citep{Xu+etal+2024}.
PSR J1741$-$0840 is reported to show the presence of nulling and subpulse
drifting.
The NF was estimated to be at $30\pm5$\% at 625 MHz \citep{Gajjar+etal+2017}.
The drifting periodicities were measured to be $4.7\pm0.6$ periods at 325 MHz
and 4.8$\pm$0.6 periods at 610 MHz \citep{Basu+etal+2016}.
\citet{Gajjar+etal+2017} concluded that the different drifting periodicities in
the trailing component is likely due to the spread in the driftband slopes.
Furthermore, none of the emission components exhibit significant changes in
driftband slope influenced by nulls, and the drifting phase memory across nulls
is not present for the trailing component \citep{Gajjar+etal+2017}.
These nulling and subpulse drifting phenomena have primarily been
investigated at low frequencies, with limited study at higher frequencies.
The low radio frequency observations are likely affected by propagation
effects through the intervening medium, such as scintillation and scattering
effects, which prompts us to carry out a detailed study on the individual pulse
emission properties of this pulsar at a higher frequency.

Here we present the thorough study on the individual pulse emission from PSR
J1741$-$0840, which show both quasiperiodic nulling and subpulse drifting.
The structure of the paper is organized as follows.
The observational setup and data reduction techniques are described in
Section~\ref{sec:obs}.
Section~\ref{sec:avg_prof} details the polarimetric behaviour of the average
pulse profile.
The single-pulse analysis leading to the characterization of subpulse drifting
and quasiperiodic nulling are present in Section~\ref{sec:sgl}.
A discussion of the implications of the results of this study is presented in
Section~\ref{sec:dis} followed by a short summary and conclusions in
Section~\ref{sec:con}.

\section{Observations and data reduction}
\label{sec:obs}

The observations on PSR J1741$-$0840 presented in this article were carried out
using the Effelsberg 100-m and Parkes 64-m radio telescopes.
At Effelsberg, the observations were performed with the central beam of the
21-cm 7-beam (P217$-$3) receiver at the central frequency of 1347.277 MHz on
2013 September 19.
The coherent dedispersion system based on the ROACH
board\footnote{https://casper.berkeley.edu/wiki/ROACH}, known as Asterix, serves
as the pulsar recording machine.
The Asterix system is particularly available to record data suitable for
high-precision timing \citep{Kramer+Champion+2013}.
It sampled over 218.75 MHz of bandwidth at 8 bits across two polarizations,
delivering baseband data which was processed in real-time to produce
coherently dedispersed pulse profiles for subsequent analysis.
The capability of the Asterix ensures the reliability and precision of the
results presented in the study.
Then online data were synchronously translated into 56 frequency channels
and 1024 phase bins in a single-polarization mode (i.e. total intensity only).
The resulted data were saved in the PSRFITS \citep{Hotan+etal+2004} format for each pulse
separately and then merged to create a single file.
Data from the Parkes observations were taken on two epochs at the central
frequency of 1369 MHz with a bandwidth of 256 MHz.
On 2014 October 17, the observations were undertaken with the central beam of
the 20-cm multibeam (MULTI) receiver and the data were recorded with the Parkes
Digital Filterbank Mark 3 (PDFB3) backend.
On 2016 April 09, the observations were carried out using the H$-$OH receiver
and the PDFB4 backend.
The MULTI system is designed with multiple beams for wide-field surveys and
high-resolution imaging across a broad frequency range \citep{Staveley+etal+1996}.
The H-OH feed is designed specifically for observing hydrogen and hydroxyl lines
with high-sensitivity for these particular frequencies \citep{Thomas+etal+1990}.
The MULTI system has challenges with beam alignment and signal purity due to
non-orthogonality and cross-polarization.
In contrast, the H-OH feed achieves a high level of performance with minimal
issues in these areas, making it ideal for precise measurements in its
specialized frequency range.
For both observations, the data were recorded at 256 $\mu$s time resolution,
spread across 512 channels, thus providing a frequency resolution of 0.5 MHz,
and were converted to single-pulse archives using the DSPSR
package\footnote{https://dspsr.sourceforge.net}
\citep{van+Bailes+2011}.
Each Parkes observation was preceded by observation of a cycled, coupled noise
diode, allowing accurate polarization calibration through measurement of differential
gain and phase.
The observation details are presented in Table~\ref{tab:obs}.

\begin{table*}
	\centering
	\caption{Observational parameters of PSR J1741$-$0840 with the Effelsberg and Parkes
	radio telescopes.}
	\begin{tabular}{ccccccccc}
		\hline\hline
		Date & Time & Telescope & Receiver & Backend & Frequency & Bandwidth & Length & Flux \\
		(yyyy$-$mm$-$ss) & (UTC) & & & & (MHz) & (MHz) & (seconds) & (mJy) \\
		\hline
		2013$-$09$-$19 & 14:35:58 & Effelsberg & P217$-$3 & Asterix  & 1347.277 & 218.750 & 7312.894 & ... \\
		2014$-$10$-$17 & 07:49:38 & Parkes & MULTI & PDFB3 & 1369 & 256 & 3605.466 & 6.33$\pm$0.97 \\
		2016$-$04$-$09 & 16:29:19 & Parkes & H$-$OH & PDFB4 & 1369 & 256 & 2165.771 & 3.75$\pm$1.08 \\
		\hline
	\end{tabular}
	\label{tab:obs}
\end{table*}

The observations were post-processed using the PSRCHIVE software
suite\footnote{https://psrchive.sourceforge.net} \citep{Hotan+etal+2004}.
The spurious radio frequency interference (RFI) signals were initially mitigated
from the raw filterbank data.
The bandpass was removed by dividing each sample in a channel by the average
value in that channel for each pulse, leading to each sample having a value
around unity.
For the narrowband RFI at a specific frequency channel, the median-filter
technique was deployed to identify and flag.
Subsequently, the frequency-dependent spread caused by the dispersive effects of
the interstellar plasma was corrected by incoherent dedispersion at the
dispersion measure of the pulsar (DM = 74.90 \pcm) \citep{Hobbs+etal+2004}.
Then, the data were folded at the predicted topocentric period using polynomial
coefﬁcients determined by TEMPO\footnote{http://tempo.sourceforge.net/} and
reorganized in the form of a pulse stack with 1024 phase bins per rotation.
The baselines are subtracted for each single pulse by removing any trends
modeled with a sixth-order polynomial as a function of pulse longitude
\citep{Lynch+etal+2013}.
Additionally, the Parkes data were flux-calibrated using the noise diode signal
that was injected into the receiver, for each polarization individually, prior
to each observation.
The polarization calibration was performed to correct for the differential gain
and phase between the two linear feed probes using the ``SingleAxis" model
\citep{Ord+etal+2004,vanStraten+etal+2012}.
The Full Stokes parameters were derotated by using the rotation measure of 124
\radm\ for every single pulse \citep{Hamilton+Lyne+1987}.

\section{Average pulse profile }
\label{sec:avg_prof}

\begin{figure}
	\centering
	\includegraphics[width=8.0cm,angle=0,scale=1]{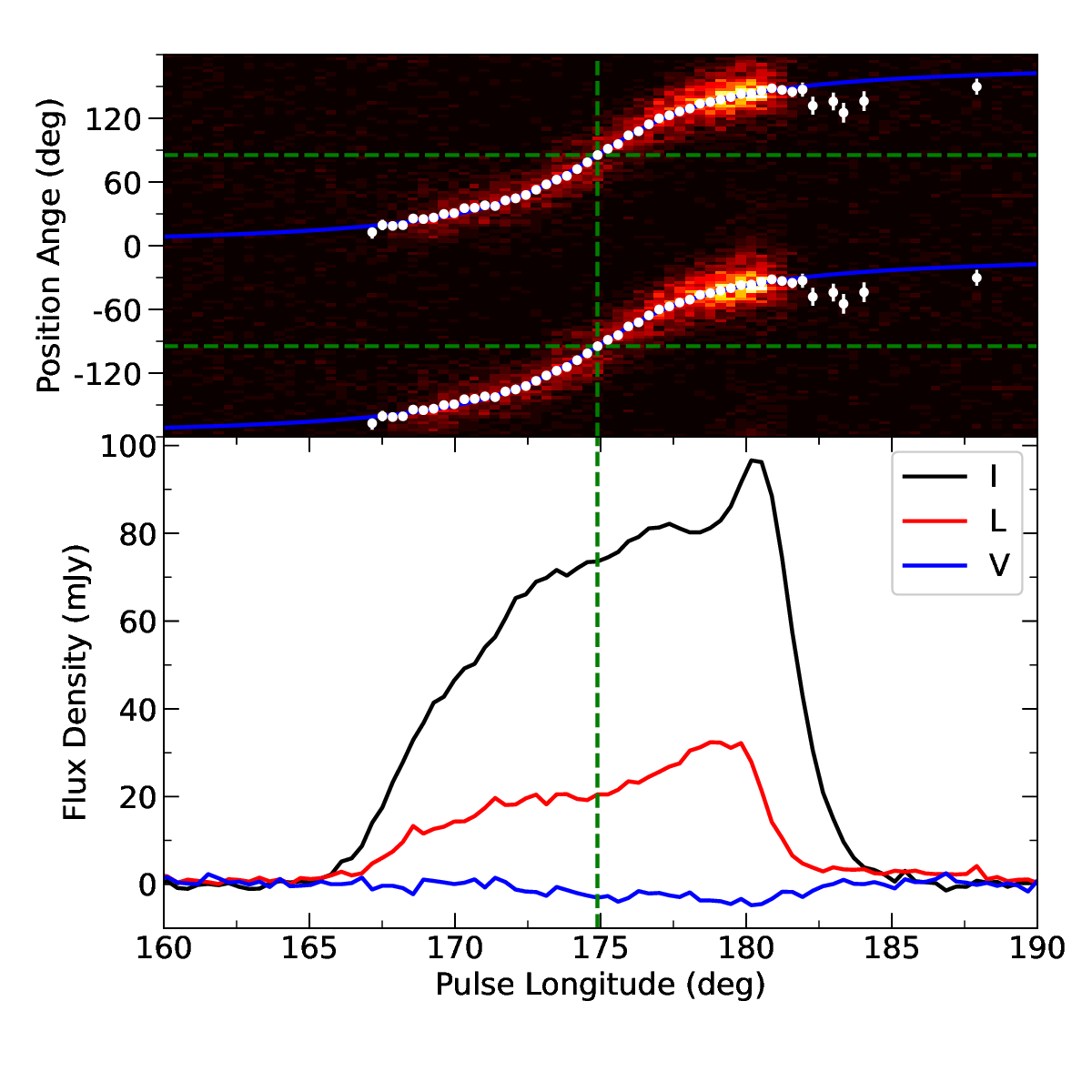}
	\caption{Integrated polarized pulse profile of PSR J1741$-$0840 computed
	from the Parkes observations at 1369 MHz.
	The lower panel gives the total intensity (Stokes $I$, solid curve), the
	total linear (Stokes $L=\sqrt{Q^2+U^2}$, red curve) and the circular
	polarization (Stokes $V$, blue curve).
	The top panel PPA [=$\frac{1}{2}\tan^{-1}(U/Q)$] histogram corresponds to
	those samples having PPA errors smaller than 10$^\circ$, and it is plotted
	twice for clarity with the average PPA traverse overplotted (white dots).
	The blue model curve is discussed in the text.}
	\label{pic:pol_profile}
\end{figure}

The complete set of Stokes parameters ($I$, $Q$, $U$, and $V$) provides a
comprehensive description of the polarization characteristics of electromagnetic
waves.
In this context, $I$ represents the overall intensity, while $Q$ and $U$ denote
the intensity of linear polarization as $L=\sqrt{Q^2+U^2}$, and $V$ 
indicates the intensity of circular polarization. 
The polarization position angle (PPA) is defined as $\psi=0.5\tan^{-1}(U/Q)$. 
The intrinsic PPA values along the pulse longitude were obtained and only
computed when the samples having PPA errors smaller than 10$^\circ$.
Figure~\ref{pic:pol_profile} shows the integrated polarimetric pulse profile for
PSR J1741$-$0840 at 1369 MHz to reveal the underlying emission mechanism and to
determine the geometry of the pulsar.
The bottom panel presents the total intensity (black), total linear polarization
(red), and circular polarization (blue).
The polarimetric pulse profile is identical to that obtained by
\citet{Johnston+Kerr+2018} with no flux calibration observed with the Parkes
radio telescope at the same frequency.
Our measurements of fractional linear polarization, fractional circular polarization
and pulse width are consistent with the values reported by \citet{Johnston+Kerr+2018}.
However, our Parkes observations are flux calibrated absolutely, and the derived
flux densities are listed in Table~\ref{tab:obs}, which were not included in
\citet{Johnston+Kerr+2018}.
It is noted that there is no paired calibrator for the Effelsberg observation as
well.
The profile is highly asymmetric and consists of multiple Gaussian components.
The circular polarization is present at low levels and appears to flip
handedness in the pulse center.
The pulsar is moderately linearly polarized and shows the presence of
substantial depolarization in the trailing edge of the profile.

Determining the geometry of radio pulsar emission is crucial for understanding
the mechanisms behind it. 
This involves primarily two angles: the inclination angle ($\alpha$), which is
the angle between the rotation axis and the magnetic axis, and the impact
angle ($\beta$), which is the angle at closet approach of the liner of sight to
the magnetic axis.
These angles allow us to determine the intrinsic beamwidth and other geometrical
properties.
Shortly after the discovery of pulsars, \citet{Radhakrishnan+Cooke+1969}
introduced the rotating vector model (RVM) to interpret pulsar polarization.
The radiation is beamed along the field lines and the plane of polarization is
determined by the angle of the magnetic field as it sweeps past the line of
sight.
For a known viewing geometry, the polarization position angle PPA ($\psi$)
corresponding to a pulse longitude ($\phi$) can be expressed as
\begin{equation}
	\psi=\psi_0+\arctan\left[\frac{\sin\alpha\sin(\phi-\phi_0)}{\sin\zeta\cos\alpha-\cos\zeta\sin\alpha\cos(\phi-\phi_0)}\right],
\end{equation}
where $\zeta=\alpha+\beta$ represents the angle between the rotation axis and the
line of sight, $\phi_0$ is the pulse longitude that corresponds to the fiducial
plane containing the magnetic and rotation axes, and $\psi_0$ is the position
angle at the point of closet approach of the observer's line of sight to the
magnetic pole.
Therefore, accurately assessing the PA traversal is crucial for constraining the
viewing geometry of the pulsar.

\begin{figure}
	\centering
	\includegraphics[width=10.0cm,angle=0,scale=0.9]{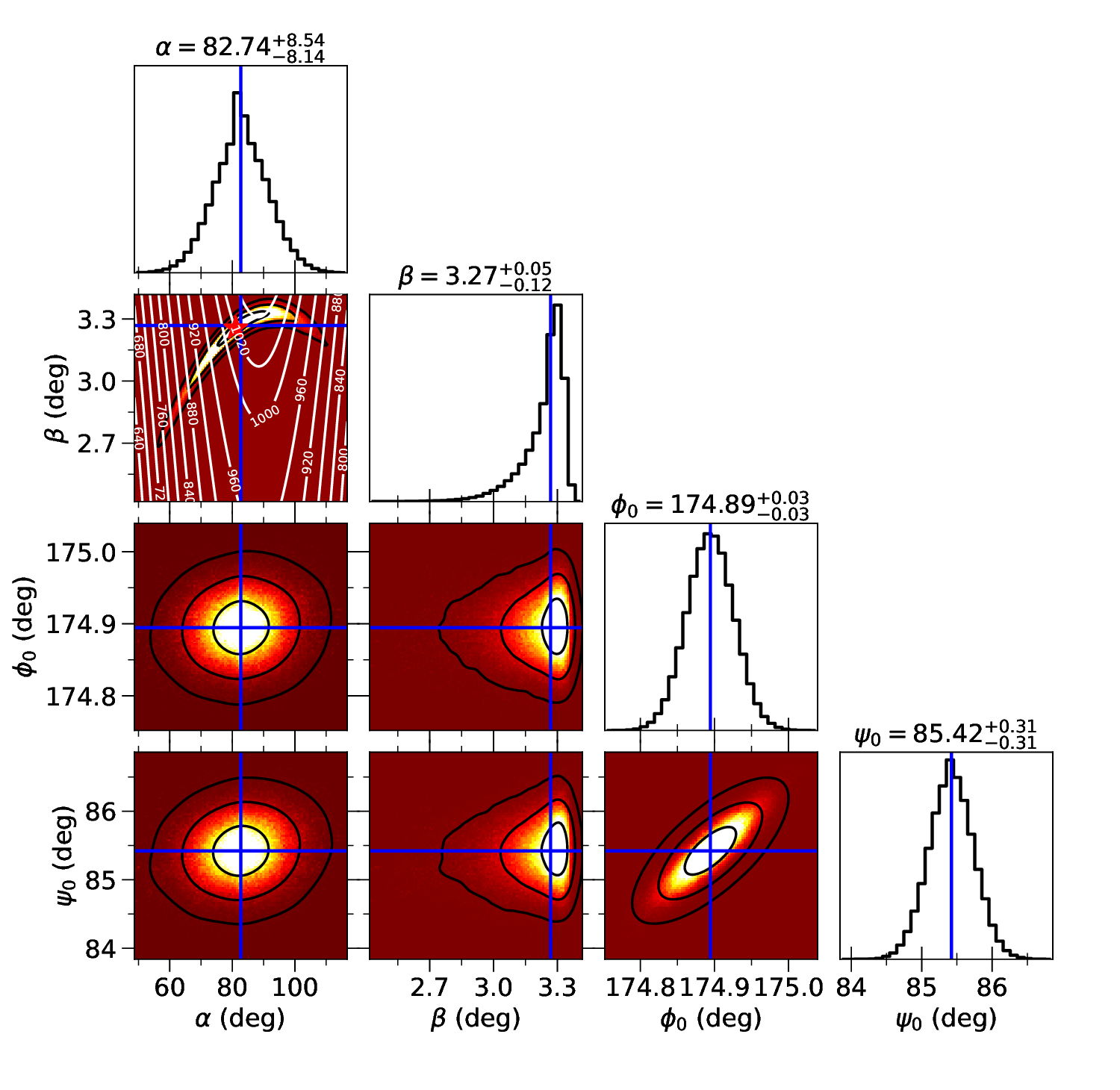}
	\caption{Posterior probability distributions for the inclination angle
	$\alpha$, impact angle $\beta$, position angle offset $\psi_0$ and the
	rotational longitude offset $\phi_0$, derived using a combination of the RVM
	model fit and a MCMC sampler.
	The top of each column illustrates the one-dimensional posterior
	distributions for each parameter, showing their probability density
	functions.
	The remaining subplots reveal the bivariate correlations between pairs of
	parameters, highlighting how they interact with each other.
	Additionally, the blue lines featured in these plots indicate the quoted
	median point estimates for each parameter, which serve as the reference
	values for the central tendency of the distributions.
	The significant covariances between $\alpha$ and $\beta$ is shown with
	superposed white contours of height in km derived by assuming a filled
	emission beam.
	The contour intersects the RVM fitting at the location of
	$\alpha=81.09^\circ$ and $\beta=3.26^\circ$, which is marked as a red
	star.
	The lower and upper uncertainties quoted for all parameters correspond
	respectively to the 16 and 84 percentiles of their probability
	distributions.
	The contours are 1$\sigma$, 2$\sigma$ and 3$\sigma$ joint confidence
	contours.
	The units of all angles is degrees.}
	\label{pic:pol_prof_corner}
\end{figure}

The upper panel of Figure~\ref{pic:pol_profile} gives the linear polarized PA
histograms for all individual pulses, as well as the average PA traversal for
PSR J1741$-$0840 at 1369 MHz.
The smooth PA traversal track existing all along with the profile shows the characteristic
S-shaped curve and can be interpreted in the frame of RVM.
In order to determine the values of $\alpha$, $\beta$, $\phi_0$ and $\psi_0$,
the Markov Chain Monte Carlo (MCMC) fits are performed by following the
technique implemented by \citet{Johnston+Kramer+2019}.
The Goodman \& Weare's Affine Invariant MCMC ensemble sampler implemented in
the python package EMCEE \citep{Foreman-Mackey+etal+2013} is adopted for
Bayesian parameter estimation.
It samples from probability distributions and estimates their properties by
constructing a Markov chain which has the desired distribution as its
equilibrium distribution.
The initial values of $\alpha$, $\beta$, $\phi_0$ and $\psi_0$ are
estimated from a least-squares fitting procedure implemented in SciPy
\citep{Virtanen+etal+2020}.
Then non-informative uniform priors are set for angles and phases between 0$^\circ$
and $360^\circ$, which offers no biases to the posterior estimates.
We initialize 32 walkers within a $\pm10\sigma$ range of the initial fit
values of these parameters.
To address the finite correlation length of the chains and generate independent
samples, we initially run a burn-in phase to eliminate the influence of initial
conditions. 
Subsequently, the parameter space are explored until each walker has
accumulated at least 1000 independent samples.
Finally, the full posterior probability distributions for the RVM model are
shown in the corner plot of Figure~\ref{pic:pol_prof_corner}.
The reported values correspond to the medians of the medians of the posterior
probability distributions, along with a $1-\sigma$ quantile estimated as half
the difference between the 16 and 84 percentiles.
It is noted that the values of $\phi_0$ and $\psi_0$ are largely
independent of $\alpha$ and $\beta$, as the inflection of the PA swing and
the value of the PA at that point are relatively model-independent.
The RVM fit to the PA is excellent with significantly constrained quantities of
$\phi_0=174.89^\circ$ and $\psi_0=85.42^\circ$,
However, it is difficult to determine the geometrical angles accurately because
$\alpha$ and $\beta$ are extremely covariant and unreliable due to the narrow
duty cycle of the pulse profile \citep{Mitra+Li+2004}.
To determine the geometry based on the shape of the PA swing, a grid search
was conducted across all possible combinations of $\alpha$ and $\beta$ to
minimize the $\chi^2$ by calculating the discrepancy between the measured and
modeled PA swing \citep{Wen+etal+2020}.
The fitted parameters include the pulse longitude of the magnetic axis and the
PA at that position.
As shown in the $\alpha-\beta$ plane, the derived `banana-shaped' $\chi^2$
contours illustrate the constraint's dependency on both parameters
simultaneously.
It is noted that the RVM fit is poorly constrained with
$2.7^\circ<\beta<3.5^\circ$ and $55^\circ<\alpha<110^\circ$.
Therefore, $\alpha$ is not well constrained by the shape of the PA swing.
And the high degree of correlation between $\alpha$ and $\beta$ complicates the
constraint on each parameter individually.
For pulsars exhibiting a pronounced slope variation across the PA swing or
a very wide longitudinal extent of the profile, the `banana' shape can be thin.
To provide a tight constraint on the $\alpha$ and $\beta$ pair, the
emission height is derived, as detailed in subsection 5.1.
The RVM corresponding to the median values is plotted as blue curve in the upper
panel of Figure~\ref{pic:pol_profile}.

\begin{figure}
	\centering
	\includegraphics[width=8.0cm,angle=0,scale=1]{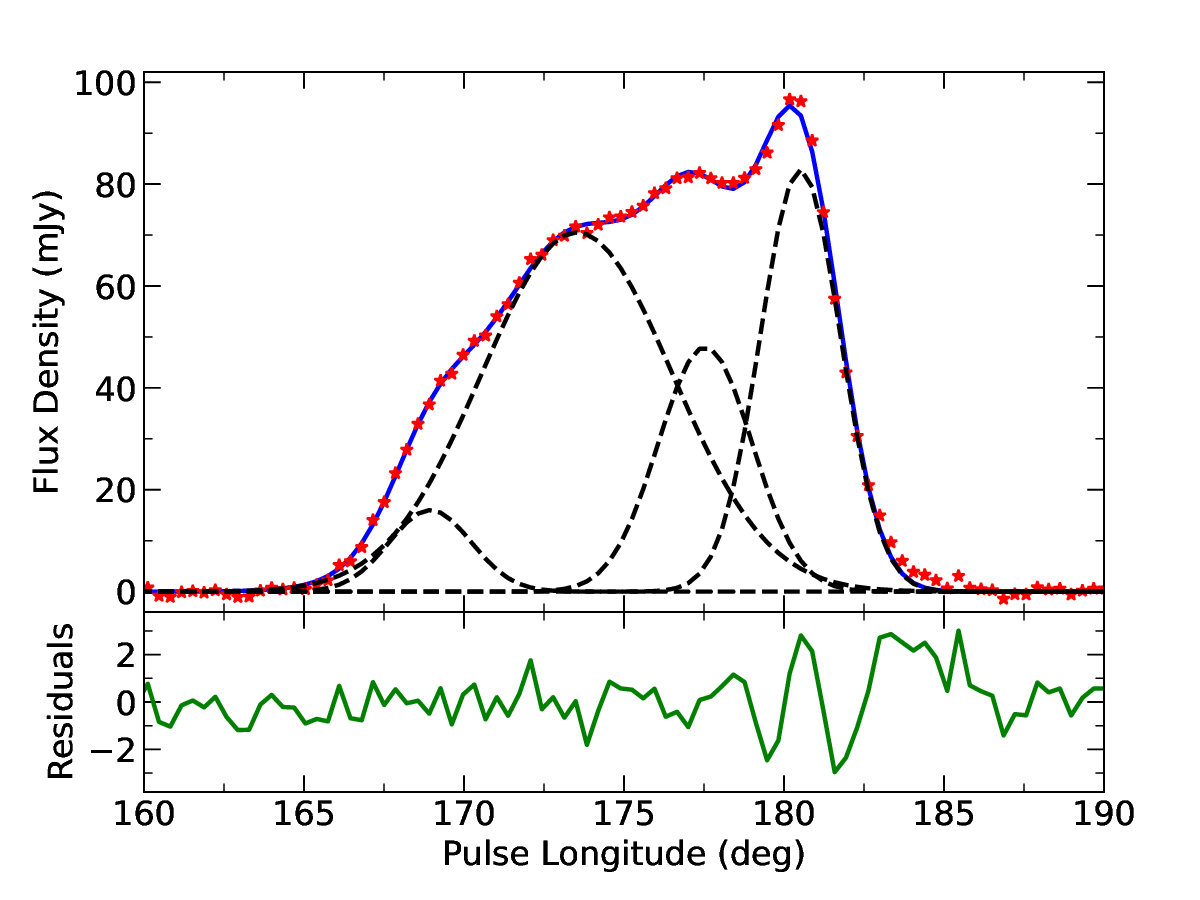}
	\caption{The observed pulse profile (red stars) for PSR J1741$-$0840 is
	shown across the on-pulse region.
	The blue solid curve represents the fit using a combination of four
	Gaussian components (black dashed lines).
	The post-fit residuals are depicted by the green curve in the lower panel.}
	\label{pic:prof_fitting}
\end{figure}

The average pulse profile is assumed to be representative of the radio emission
region within the pulsar magnetosphere, which can be separated into individual
components.
Following the technique described by \citep{Kramer+1994}, each profile component
is assumed to have a Gaussian shape, and the integrated pulse profile could be
fitted with a multi-Gaussian function
\begin{equation}
	S(\phi) = \sum_{i=1}^{n}S_{\rm{peak},\it{i}}\exp\lbrack-4\ln2(\frac{\phi-\phi_i}{\it{W_{\rm{50},\it{i}}}})^2],
\end{equation}
where $S_{\rm{peak},\it{i}}$, $\phi_i$, and $W_{50,i}$ are the peak flux,
position of peak, and full width at half maximum (FWHM) of the $i$th component,
respectively.
We calculate the post-fit residuals within the fitted on-pulse region and
compare their distribution to that of the off-pulse region.
The different statistical tests are then applied to compare the two residual
distributions, including the F-test, Students-T-test, and the non-parametric
Kolmogorov-Smirnov-test.
All tests yield the probabilities greater than the significance level of 5\%
until the fourth component is added to the sum.
As a result, we cannot reject the null-hypothesis that the post-fit-residuals in the
on-pulse and off-pulse regions are drawn from the same statistical
parent-distribution.
Figure~\ref{pic:prof_fitting} shows the results of fitting Gaussians to the
average pulse profile of PSR J1741$-$0840.
The sum of four Gaussian components is superposed on the data.
The post-fit residuals in the on-pulse region show a noise-like pattern with a
distribution that is similar to the distribution of the off-pulse data.
The best-fitted parameters determined using the Levenberg–Marquardt gradient
algorithm are presented in Table~\ref{tab:prof_param}.
As suggested by \citet{Basu+etal+2021}, the pulsar's profile can be
interpreted as a sightline traverse cutting across two concentric rings of inner
and outer cones.

\begin{table}
	\centering
	\caption{Parameters of the fitted profile components for PSR J1741$-$0840.}
	\begin{tabular}{cccc}
		\hline\hline
		Component & $S_{\rm{peak}}$ & $\phi$ & $W_{50}$ \\
		 & (mJy) & (deg) & (deg) \\
		 \hline
		 C1 & 16.00$\pm$2.93 & 168.95$\pm$0.09 & 3.02$\pm$0.35 \\
		 C2 & 70.51$\pm$1.09 & 173.53$\pm$0.17 & 7.02$\pm$0.48 \\
		 C3 & 48.03$\pm$5.70 & 177.54$\pm$0.11 & 3.46$\pm$0.30 \\
		 C4 & 82.90$\pm$2.85 & 180.51$\pm$0.06 & 2.95$\pm$0.07 \\
		\hline
	\end{tabular}
	\label{tab:prof_param}
\end{table}

\section{Single pulse sequence}
\label{sec:sgl}
\begin{figure}
	\centering
	\includegraphics[width=8.0cm,angle=0,scale=1]{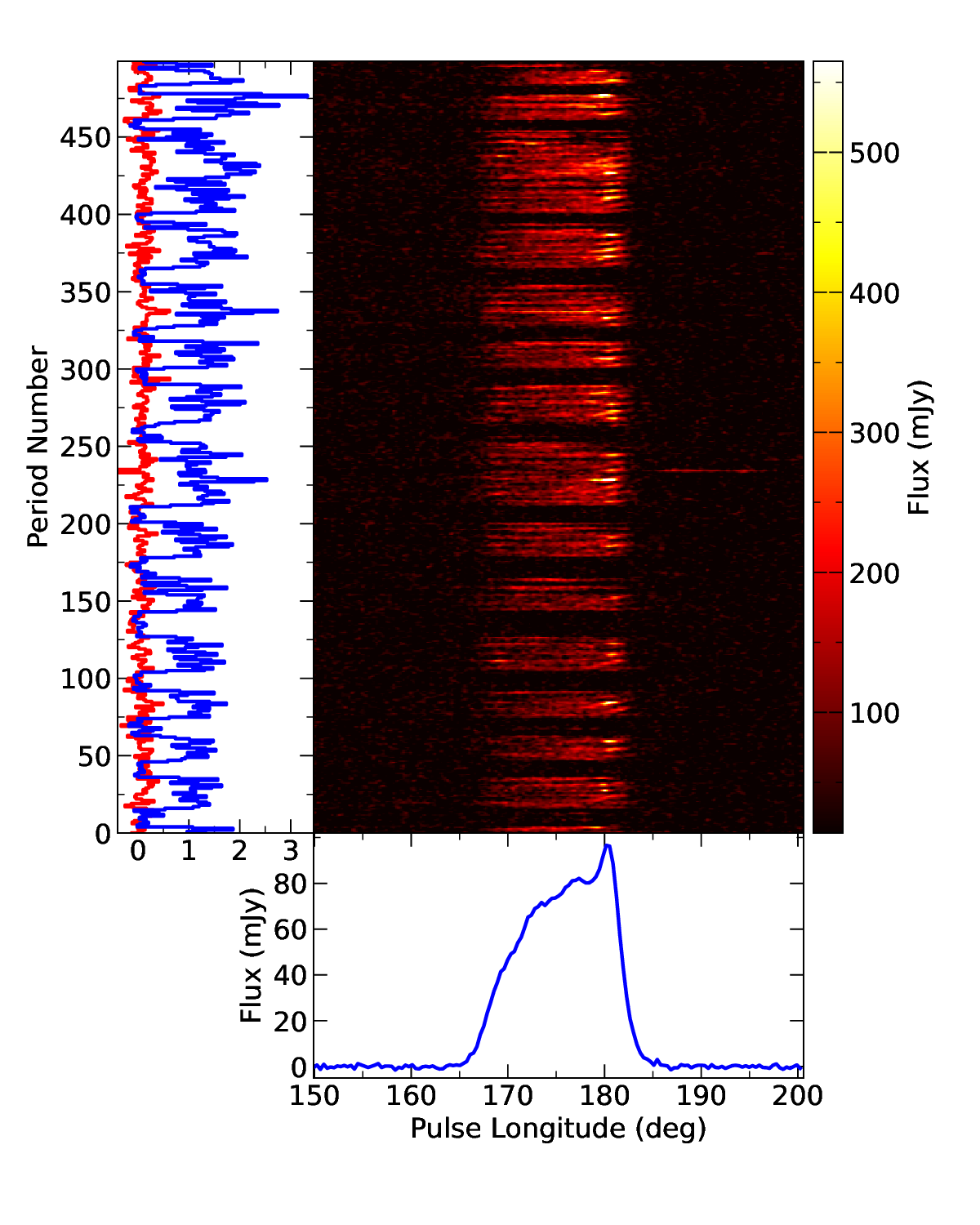}
	\caption{Top panel: a segment of a single-pulse sequence consisting of 500
	consecutive periods for PSR J1741$-$0840, observed with the Effelsberg 100 m radio
	telescope at 1347 MHz.
	Time progresses from left to right across the pulse window and from bottom
	to top with pulse number.
	Left panel: energy sequences for the on-pulse (blue) and off-pulse (red) 
	averaged across all longitude bins inside and outside the on-pulse window.
	Bottom panel: the integrated pulse profile from the entire observations.}
	\label{pic:sgl_demo}
\end{figure}

Single pulses preserve specific information regarding on the pulsar radio
emission and propagation in the magnetosphere during each individual rotation.
Exploring the individual pulse emission behaviour can provide insights into the
underlying mechanisms governing pulse emission.
Figure~\ref{pic:sgl_demo} shows a section of the color-coded individual-pulse
sequence from PSR J1741$-$0840 in the main panel, which reveals the presence of
nulling behaviour.
The pulsar goes to the null state where the radio emission vanishes.

\subsection{Pulse energy distribution}
The investigation of the pulse energy distribution provides insights into the
radio pulsar emission mechanism and the physical states of pulsar
magnetospheres.
In order to characterize the nulling behaviour of the pulsar, the on-pulse and
off-pulse energies are calculated by integrating the flux densities over the
phase ranges with equal width defined by the on-pulse and off-regions, respectively.
The relative pulse energies are then obtained by scaling with the average value
of the on-pulse energy.
An extract of the time series of the integrated flux estimated from on-pulse
(blue) and off-pulse (red) regions are shown in the left panel of
Figure~\ref{pic:sgl_demo}.
During the null state, the on-pulse energy decreases to the equivalent level to
the off-pulse energy.
Additionally, a quasiperiodic behaviour of nulling is shown.

\begin{figure*}
	\centering
	\includegraphics[width=9.0cm,angle=0,scale=0.98]{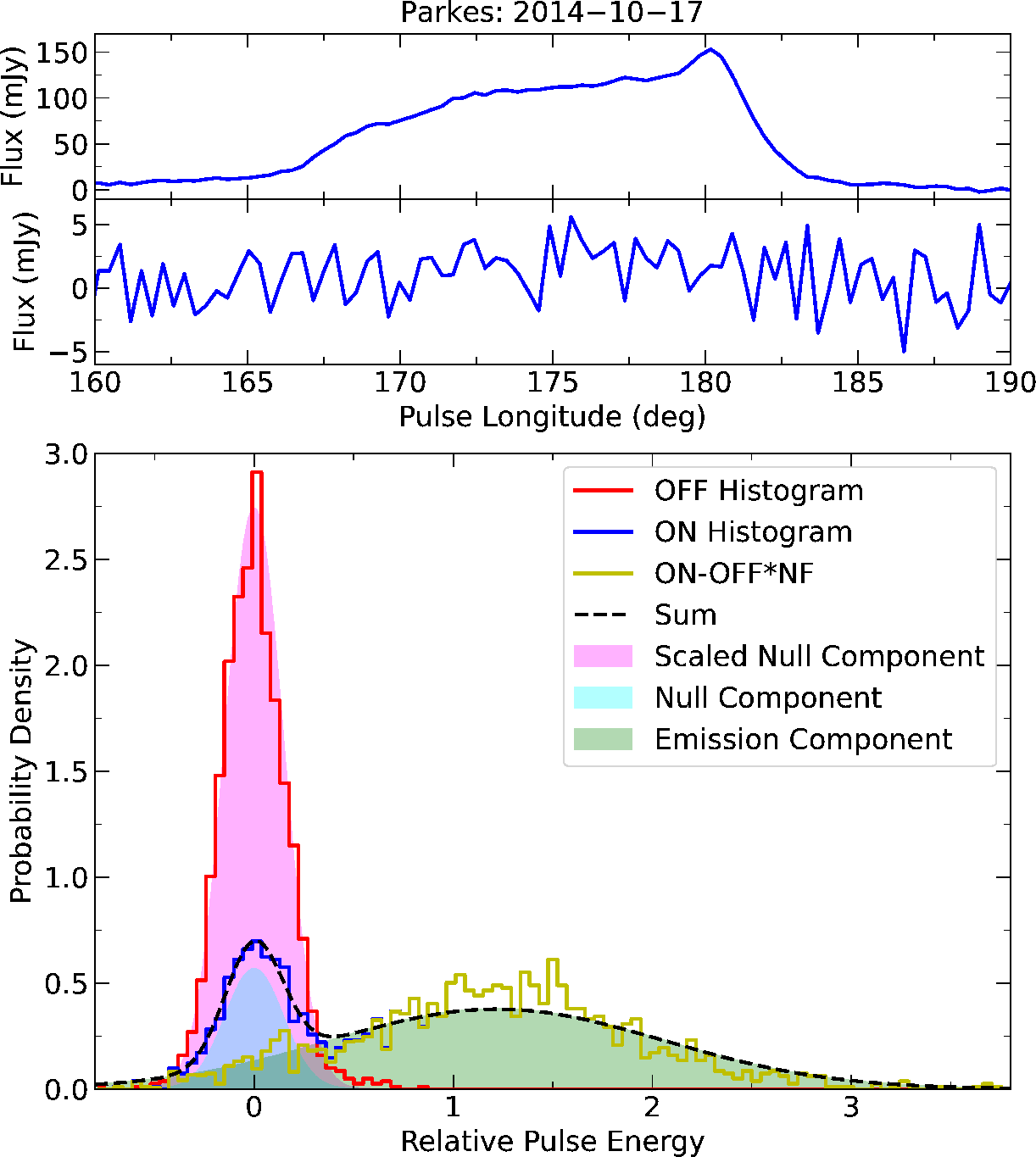}
	\includegraphics[width=9.0cm,angle=0,scale=0.98]{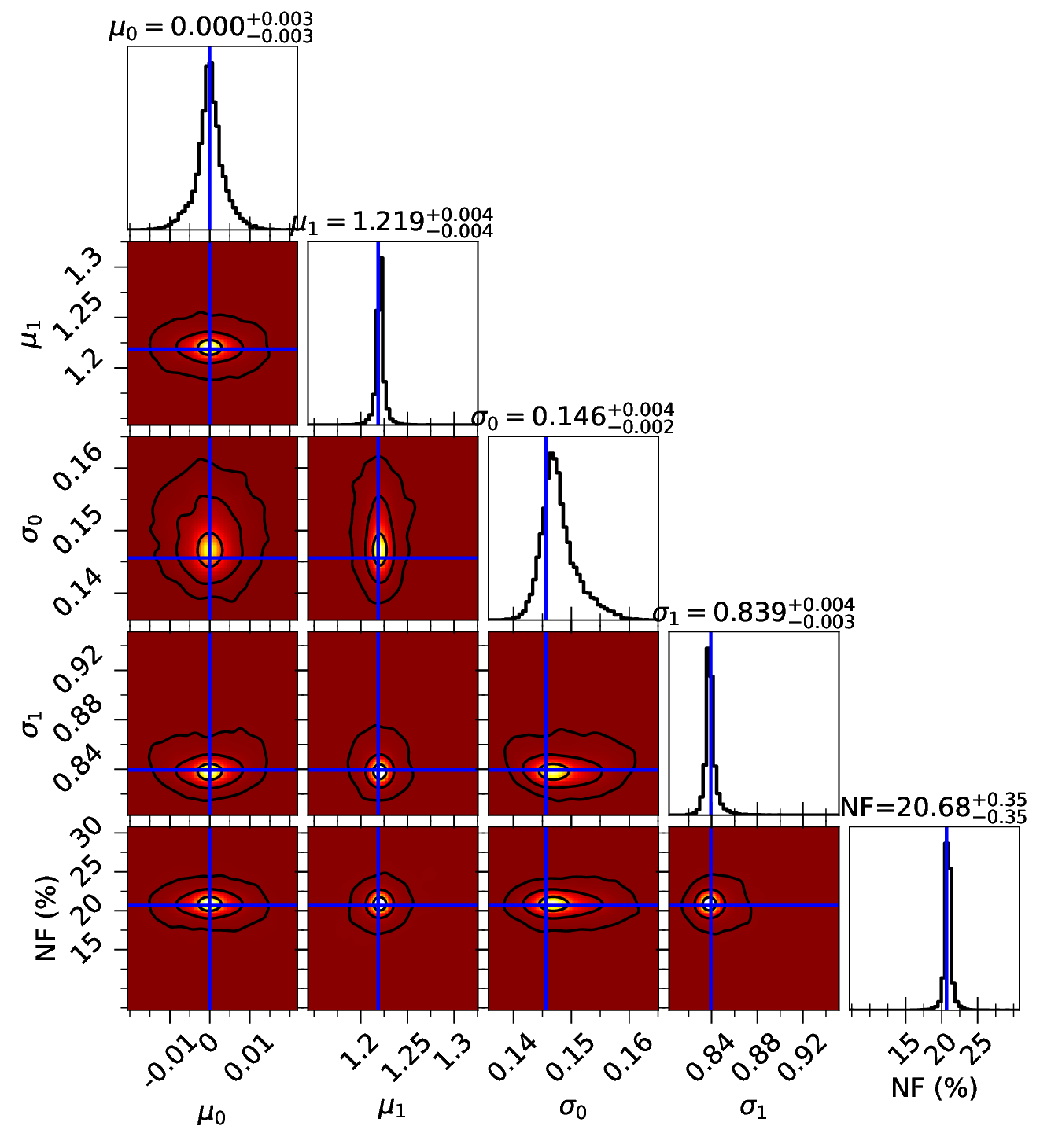}
	\caption{The top left panel presents the integrated pulse profiles obtained
	separately from burst (upper) and null (lower) pulses.
	The null profile does not show any detectable profile component, indicating
	the absence of any weak level emission.
	The bottom left panel displays the relative pulse energy distributions
	for the on-pulse window (blue) and off-pulse window (red), with
	baseline-subtracted energies normalized to the mean on-pulse energy.
	The cyan- and green-filled regions show the fit for the null and emission
	components, respectively.
	The magenta-filled region presents the component for the nulls scaled by
	1/NF, which matches the off-pulse energies well.
	The black dotted line shows the overall fit for the on-pulse distribution.
	The solid yellow line shows the on-pulse histogram after subtraction of the
	null pulses.
	The right column presents the posterior probability densities for the
	Gaussian-distribution parameters of the null and emission components derived
	using the MCMC algorithm.
	The top of each column shows the one-dimensional posterior for each
	parameter, while the other plots show the corresponding correlations between
	them.
	The solid blue lines indicate the median values with lower and upper
	uncertainties corresponding to the 16 and 84 percentiles of their
	probability distributions.
	The contours are 1$\sigma$, 2$\sigma$ and 3$\sigma$ joint confidence
	contours.}
	\label{pic:ene_dist_pks141017}
\end{figure*}

The left lower panel of Figure~\ref{pic:ene_dist_pks141017} shows the
baseline-subtracted relative pulse energy distributions for the on-pulse (blue) and
off-pulse (red) regions, in which the emission component is easily discernible from
the noise.
The off-pulse histogram can be accurately described by a single Gaussian
component centred around zero, which is attribute to the radiometer noise
entirely.
While the on-pulse histogram has two Gaussian components corresponding to the
radiometer noise (or null component) and pulsar bursting component.
Previously, the majority of studies
\citep{Wang+etal+2007,Wen+etal+2016,Wen+etal+2022} estimate NF using the
methodology proposed by \citet{Ritchings+1976}, which give biased estimation of
NF, especially for weak pulsars observed by less sensitive telescopes.
\citet{Gajjar+etal+2017} determined the NF of PSR J1741$-$0840 to be $30\pm5$\%
at 625 MHz by minimizing the expression ON$-$OFF$\times$NF for energies $\le$0
through trailing several values of NF.
The lack of a clear distinction between on-pulse and off-pulse histograms leads
to an overestimation of the NF.
And the standard method assumes that the pulses with pulse energy less than
0 are entirely due to nulling, excluding weak pulses overwhelmed by radiometer
noise or RFI, which causes the overestimation of the NF as well.
However, the values of NF are underestimated to be around 16\% at 333 and
618 MHz observed by \citet{Basu+etal+2017}. 
Even though the null pulses and burst pulses are well-resolved in the average
pulse energy distributions, the underestimation of the obtained NF maybe
resulted from the arbitrary selection of binning while constructing energy
histograms.
In this paper, we adopt an alternate method proposed by \citet{Kaplan+etal+2018}
to model the pulse energies and estimate the NF.
The likelihood of the on-pulse energies is defined as a one-dimensional
Gaussian mixture model (GMM) with two components, which corresponds to the standard
description of pulsars with nulling and bursting states.
Such a GMM probability distribution is parameterized by the means ($\mu_0$,
$\mu_1$), standard deviations ($\sigma_0$, $\sigma_1$), and weights ($\omega_0$,
$\omega_1$), with the constraint that $\omega_0+\omega_1=1$. 
The Gaussian component having the lowest mean describes null pulses, so NF is
equal to $\omega_0$.
The GMM is utilized to derive an initial fit for the on-pulse energy histogram
by using the expectation-maximization algorithm. 
The reasonable results are obtained showing significant separation between nulling
and bursting pulses.
Then the null-component parameters ($\mu_0$, $\sigma_1$) are constrained further
by way of the off-pulse likelihood function with a single Gaussian component.
The derived prior distributions are provided for a more precise fitting for the
null and bursting components of the on-pulse energy histogram by conducting a
MCMC analysis.
The details of exploring the parameter space are implemented by using the
program pulsar\_nulling\footnote{https://github.com/dlakaplan/nulling-pulsars}
developed by \citet{Anumarlapudi+etal+2023}.
The EMCEE ensemble sampler is used to sample the posterior.
We initialize 40 walkers around the best-fit region for the parameters as
determined by expectation-maximization run on the on-pulse energies.
To achieve burn in, the walkers are then run through 50 iterations to erase the
initial conditions.
The final population representative of the parameter's joint posterior is
obtained by running the walkers for 1000 iterations.
As shown in the lower left panel of Figure~\ref{pic:ene_dist_pks141017}, the
magenta-, cyan- and green-filled regions show the fit for the scaled null,
null, and bursting components, respectively.
It is noted that the scaled null component is close to the off-pulse histogram.
The off-pulse energy histogram represents the telescope noise or a Gaussian
random noise, the width of which is generally used to estimate the rms
fluctuations of the data.
The presence of nulling can be seen for pulses near the zero pulse energy as
shown in Figure~\ref{pic:sgl_demo}.
Therefore, during the nulling state, the on-pulse energy also represents the
telescope noise similar to the off-pulse energy.
The height of the zero centered energy in the on-pulse histogram provides the
fraction of the pulses in the null state, which is similar to the off-pulse
histogram.
Therefore, a scaled version of the off-pulse histogram can be well modelled for
the on-pulse histogram to estimate the NF.
The black dotted curve shows the overall fit for the on-pulse histogram.
The posterior probability densities for the Gaussian-distribution parameters of
the null and emission components are present in the right column of
Figure~\ref{pic:ene_dist_pks141017} with the point estimates of the NF from GMM
given in Table~\ref{tab:nulling_param}.
The variation of NF is likely due to the different null sequences existing in
the three observations with different length.
The average value of NF is determined to be 23$\pm$2\%.

\begin{table*}
	\centering
	\caption{Nulling properties of PSR J1741$-$0840}
	\begin{tabular}{ccccc}
		\hline\hline
		Date & NF & Nulling Periodicity & Null length & Burst Length \\
		(yyyy$-$mm$-$ss) & (\%) & (periods) & (periods) & (periods) \\
		\hline
		2013$-$09$-$19 & $23.73^{+0.17}_{-0.16}$ & $31.35^{+5.43}_{-4.03}$ & $2.09^{+0.16}_{-0.16}$ & $10.24^{+0.67}_{-0.68}$ \\
		2014$-$10$-$17 & $20.68^{+0.35}_{-0.35}$ & $32.19^{+3.03}_{-2.55}$ & $1.79^{+0.22}_{-0.22}$ & $7.35^{+0.77}_{-0.74}$ \\
		2016$-$04$-$09 & $25.53^{+0.20}_{-0.19}$ & $33.35^{+1.07}_{-1.00}$ & $1.19^{+0.15}_{-0.15}$ & $5.61^{+0.28}_{-1.06}$ \\
		\hline
	\end{tabular}
	\label{tab:nulling_param}
\end{table*}

\begin{figure*}
	\centering
	\includegraphics[width=18.0cm,angle=0,scale=1]{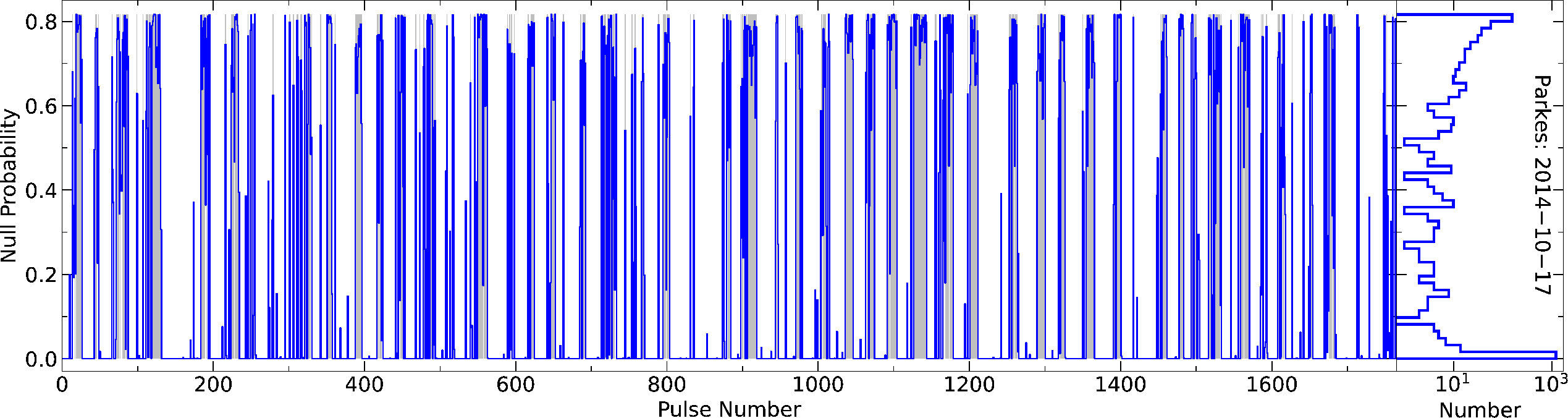}
	\caption{The left panel shows the null probabilities for every single pulse.
	For clarity, the identified null states are indicated with grey bars.
	The right panel presents the distribution of null probability, where the
	evidence for null and burst pulses is clear.}
	\label{pic:null_probability_pks141017}
\end{figure*}

In order to explicitly evaluate the probability of a given single pulse being a
null, the nulling responsibility for each individual pulse is calculated by
following the equation given by \citet{Kaplan+etal+2018}
\begin{equation}
	P = \frac{\omega_0 \mathcal{N}(\mu_0,\sigma_0)}{\sum_{i=0}^1\omega_i \mathcal{N}(\mu_i,\sigma_i)},
\end{equation}
where $\mathcal{N}(\mu_i,\sigma_i)$ represents the probability densities of the
individual pulses for the $i$th Gaussian component.
The nulling responsibility represents how likely a single pulse being a null.
The left panel of Figure~\ref{pic:null_probability_pks141017} shows the null
probabilities for every single pulse observed on 17 October, 2014 using the
Parkes telescope.
A threshold of 0.5 is generally used to divide all individual pulses into
two classes \citep{Anumarlapudi+etal+2023}.
To justify using 0.5 as the threshold, all the single pulses are arranged in the
descending order of their nulling responsibility.
A box is selected which encloses all the significant burst pulses at the low
nulling responsibility end.
To include most of the weak burst pulses, the lower end of the box is moved from
the low responsibility end towards the high responsibility end until the pulses
outside the box do not show a significant profile.
The derived boundary responsibility is close to 0.5.
Therefore, the single pulses with null probability $>$0.5 are marked as nulls, and the
pulses below the threshold are identified as burst pulses.
For clarity, the identified null states are indicated with grey bars.
The distribution of null probability is shown in the right panel, where the
evidence for two classes of pulses is clear.
The upper left panel of Figure~\ref{pic:ene_dist_pks141017} confirms that these
nulls clearly show the absence of any detectable emission for all components in
PSR J1741$-$0840.

\subsection{Quasiperiodic nulling}

\begin{figure}
	\centering
	\includegraphics[width=8.0cm,angle=0,scale=1]{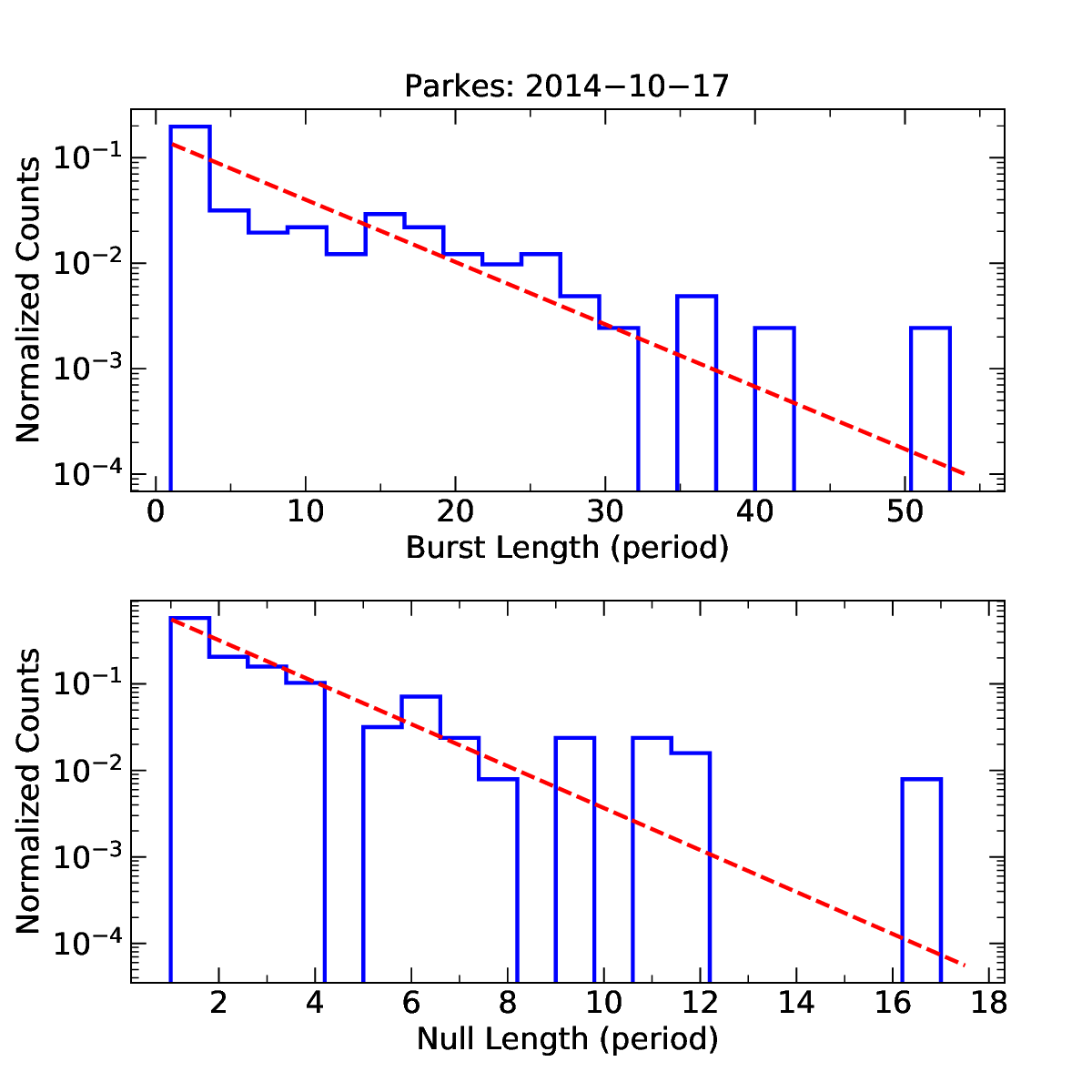}
	\caption{Duration distributions for the null (bottom) and burst (top)
	emission states	for PSR J1741$-$0840 (blue curve) is shown along with the best fit
	with an exponential model (red dashed line).
	The distributions are shown with the measured counts on a log-scale to bring
	out the details for longer pulses.}
	\label{pic:state_length_pks141017}
\end{figure}

To comprehensively understand the nulling phenomenon and determine the
clustering feature of bursting and nulling pulses,
\citet{Sheikh+MacDonald+2021} proposed that additional parameters, such as
null lengths and null randomness are required.
The durations of contiguous emission states are measured and designated as burst
length, while the intervals between these bursts are referred to as
null length.
The duration distributions for both emission and null states are
illustrated in Figure~\ref{pic:state_length_pks141017}, showing the occurrence
of the states decreases at longer timescales.
It appears that the durations of nulls and bursts can be modeled with an
exponential distribution \citep{Gajjar+etal+2012}.
The characteristic timescales for null and burst states are derived through
least-square fitting and tabulated in Table~\ref{tab:nulling_param}.
On 2014 October 17, the emission episodes have a characteristic length of 
approximately 7 periods, while the nulls last about 2 periods,
which is approximately consistent with the estimated NF of 21\%.
It is noted that the observations at three epochs exhibit distinct timescales
for null and burst states.

\begin{figure}
	\centering
	\includegraphics[width=8.0cm,angle=0,scale=1]{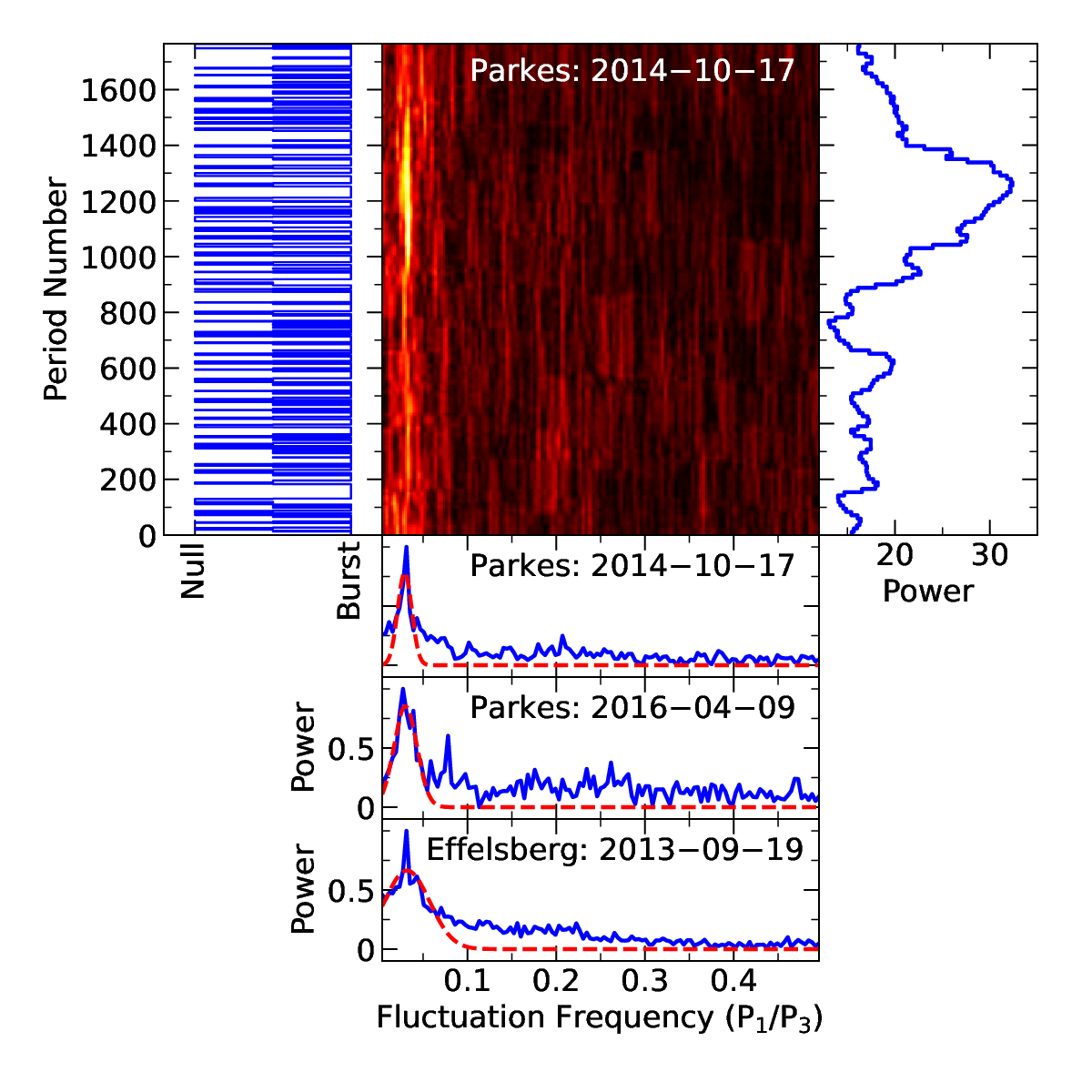}
	\caption{Left panel: the binary sequence of null and burst pulses of PSR
	J1741$-$0840 for the data obtained on 2014 October 17.
	Main panel: the temporal evolution of the fast Fourier transform of the
	null/burst time series.
	Right panel: the time-dependent fluctuation power spectra.
	Lower panels: the overall fluctuation spectra over the entire sequence at
	three epochs.
	A normal distribution is fit to the fluctuation power that is caused by the
	quasiperiodicity in the nulling, shown in red dashed curve.}
	\label{pic:quasiperiodic_nulling}
\end{figure}

In order to further explore whether the locations and durations of nulls
in PSR J1741$-$0840 are completely random or exhibit quasiperiodicity, a
one-dimensional discrete Fourier transform (DFT) is carried out on the binary
time series representing nulls and burst pulses \citep{Wang+etal+2023}.
Generally, the number of 256 consecutive points is utilized to carry out the DFT.
The initial period is then shifted by 10 pulses in successive steps until
the end of the observation period.
This technique guarantees that all subpulse information is eliminated, 
enabling the investigation of periodicity in the transitions between the burst
and null states.
Figure~\ref{pic:quasiperiodic_nulling} presents the time-varying Fourier
transform of the null and burst sequence for PSR J1741$-$0840.
The quasiperiodic nulling behaviour is seen as evidenced by a low-frequency
feature at $\sim$0.03 cycles per period (cpp) in the ﬂuctuation spectra.
The corresponding periodicities at three epochs are listed in
Table~\ref{tab:nulling_param}, where the uncertainties are determined as the
full width at half maximum of the modulation feature.
The average value of nulling periodicity is determined to be 32.30$\pm$0.82
periods.
Furthermore, it is noted that the quasiperiodic power manifests as significant
temporal variation, as depicted in the right panel.

\subsection{Subpulse drifting}

In order to explore the subpulse modulation properties of PSR J1741$-$0840,
the general techniques based on Fourier transform are utilized, including the
longitude-resolved fluctuation spectra \citep[LRFS,][]{Backer+1970}, the
harmonic-resolved fluctuation spectra \citep[HRFS,][]{Deshpande+Rankin+2001} and
the two-dimensional fluctuation spectra \citep[2DFS,][]{Edwards+Stappers+2002}.

\begin{figure}
	\centering
	\includegraphics[width=8.0cm,angle=0,scale=1]{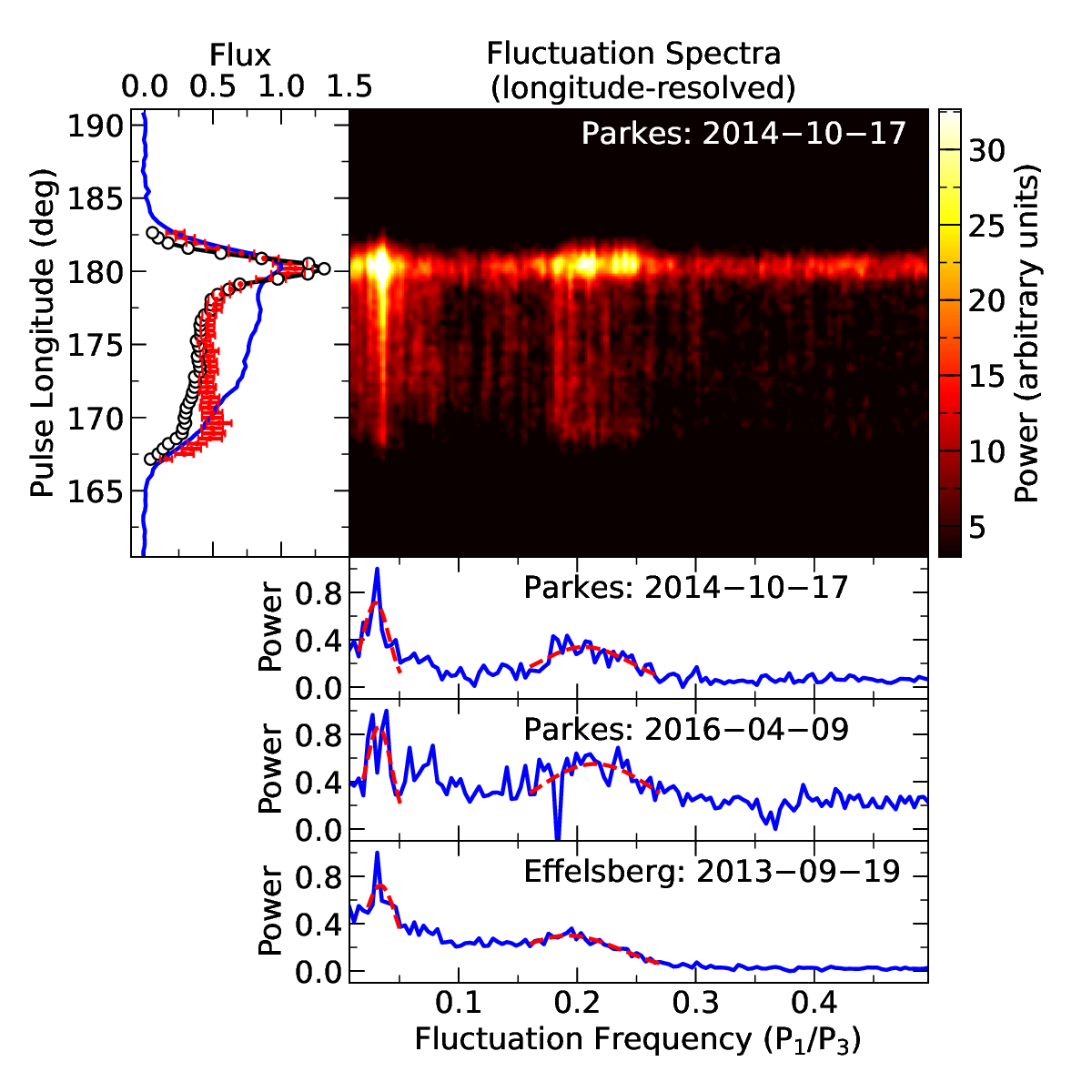}
	\caption{The longitude-resolved fluctuation power spectra for PSR J1741$-$0840
	based on the data from 2014 October 17, reveal distinct periodic modulation.
	The abscissa represents the frequency of the drifting periodicity, while the
	ordinate denotes the pulse longitude in degrees.
	A 256-point Fourier transform, averaged across the blocks of the entire
	pulse sequence, is employed.
	The intensity of drifting periodicities as a function of pulse longitude
	is depicted using color contours.
	In the left panel, the integrated pulse profile is shown by the solid blue
	curve, with the longitude-resolved modulation index indicated by red points
	with error bars, and the longitude-resolved standard deviation represented
	by black open circles.
	The lower panels display the power density spectra, vertically integrated 
	from the longitude-resolved fluctuation spectra at three different epochs.
	The Gaussian fits to the modulation features are overlaid in red.}
	\label{pic:lrfs}
\end{figure}

The main panel in Figure~\ref{pic:lrfs} displays the spectral power of ﬂuctuations
as a function of rotational longitude for data collected on 2014 October
17, which reveals distinct periodic modulation features across the 
entire emission window.
The left panel shows the integrated pulse profile of PSR J1741$-$0840,
normalized to the peak intensity and indicated with the blue solid curve.
The longitude-resolved modulation index and the longitude-resolved standard
deviation are shown with the red points with error bars and the black dotted
curve, respectively.
The modulation indices are measured to be around 0.5 at the centre of components
with high modulation index at the trailing component, which indicates that the
variation of pulse intensity at the trailing component is significant.
The lower panels present the integrated power spectra over the on-pulse window
for three observations.
Alongside a low-frequency ($\sim$0.03 cpp) feature associated with
the periodic nulling, a high-frequency ($\sim$0.2 cpp) modulation feature is
presented as well.
Following the method outlined by \citep{Chen+etal+2024}, the fluctuation
periodicities and their corresponding uncertainties are determined by fitting a
Gaussian to the spectral power of these features.
The derived modulation periodicities ($P_3$) from LRFS are tabulated in
Table~\ref{tab:drifting_param}.

\begin{figure}
	\centering
	\includegraphics[width=8.0cm,angle=0,scale=1]{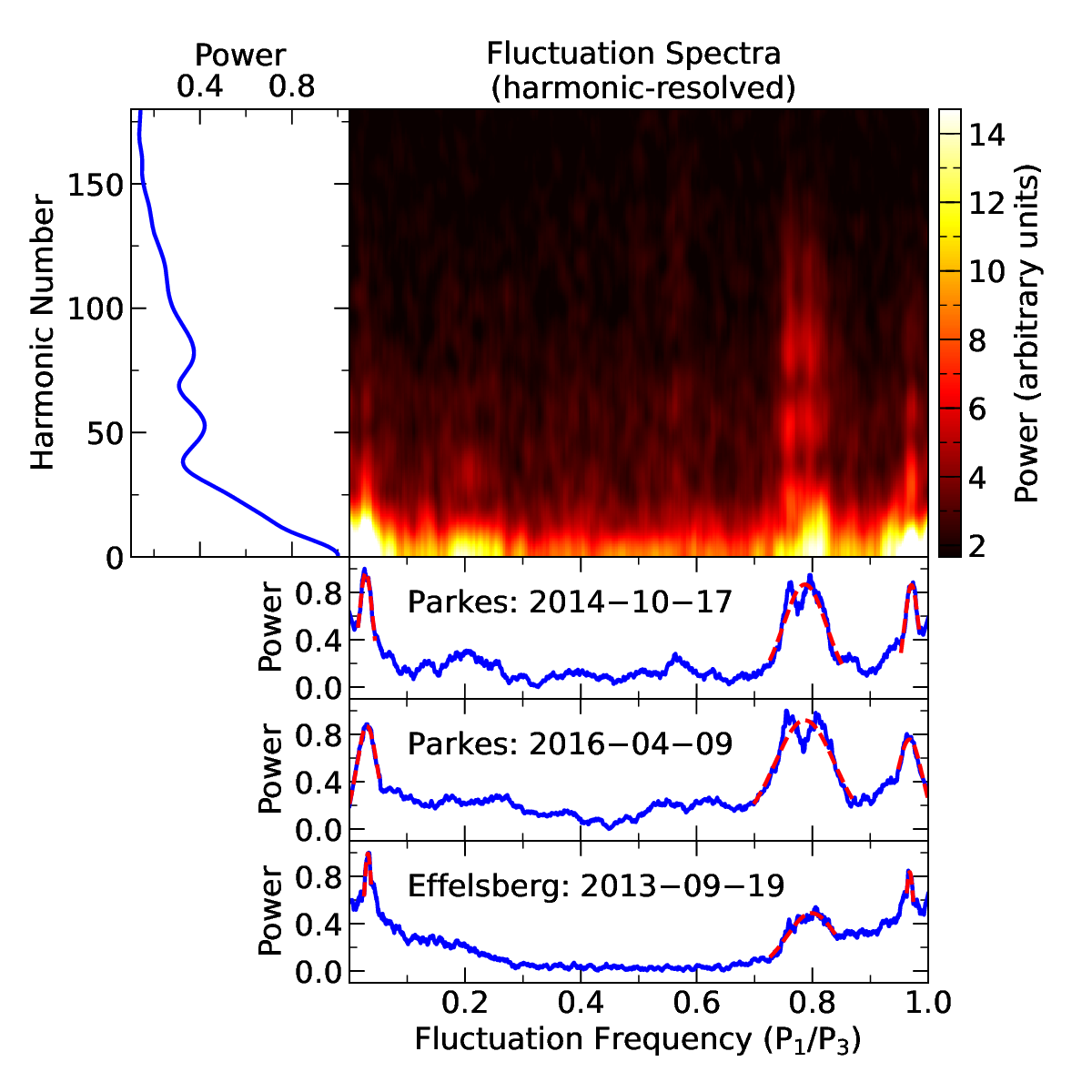}
	\caption{The harmonic-resolved fluctuation power spectra for PSR
	J1741$-$0840 are presented as a color-coded plot in the main panel, with the
	lower panels showing the harmonic-integrated spectra across three epochs.
	Three distinct modulation features are evident: a low-frequency feature near
	0.03 cpp and its symmetric counterpart around 0.97 cpp, alongside a
	high-frequency feature appearing at approximately 0.82 cpp. 
	The left panel displays the amplitude of the harmonics at the integral multiple
	of the pulsar's rotation frequency.}
	\label{pic:hrfs}
\end{figure}

In order to address the aliasing issues caused by the poor sampling of
ﬂuctuations at a given longitude only once every pulse period, the HRFS is
calculated to explore the phase behaviour of the fluctuation features.
The main panel in Figure~\ref{pic:hrfs} presents the computed HRFS for PSR
J1741$-$0840 from observations on 2014 October 17.
The left panel shows the spectral power at a frequency of 1 cycle per period and
its harmonics.
The bottom panels show the integrated fluctuation spectra across three epochs.
The low-frequency fluctuation is predominantly amplitude-modulated drifting, as
shown by its symmetrical HRFS.
This symmetry suggests that the subpulses do not traverse across
the pulse window, but instead exhibit periodic changes in intensity, 
indicative of the feature of quasiperiodic nulling.
Conversely, the high-frequency ﬂuctuation, as seen in the LRFS,
appears in the HRFS with the corresponding peak shifted to $\sim$0.8 cpp, 
reflecting a gradual shift of the subpulses toward the trailing edge.
Additionally, it is noted that the fluctuation power due to subpulse drifting
reaches its maximum around harmonic numbers of 80, suggesting that $P_2$
is approximately $4.5^\circ$.

\begin{figure}
	\centering
	\includegraphics[width=8.0cm,angle=0,scale=1]{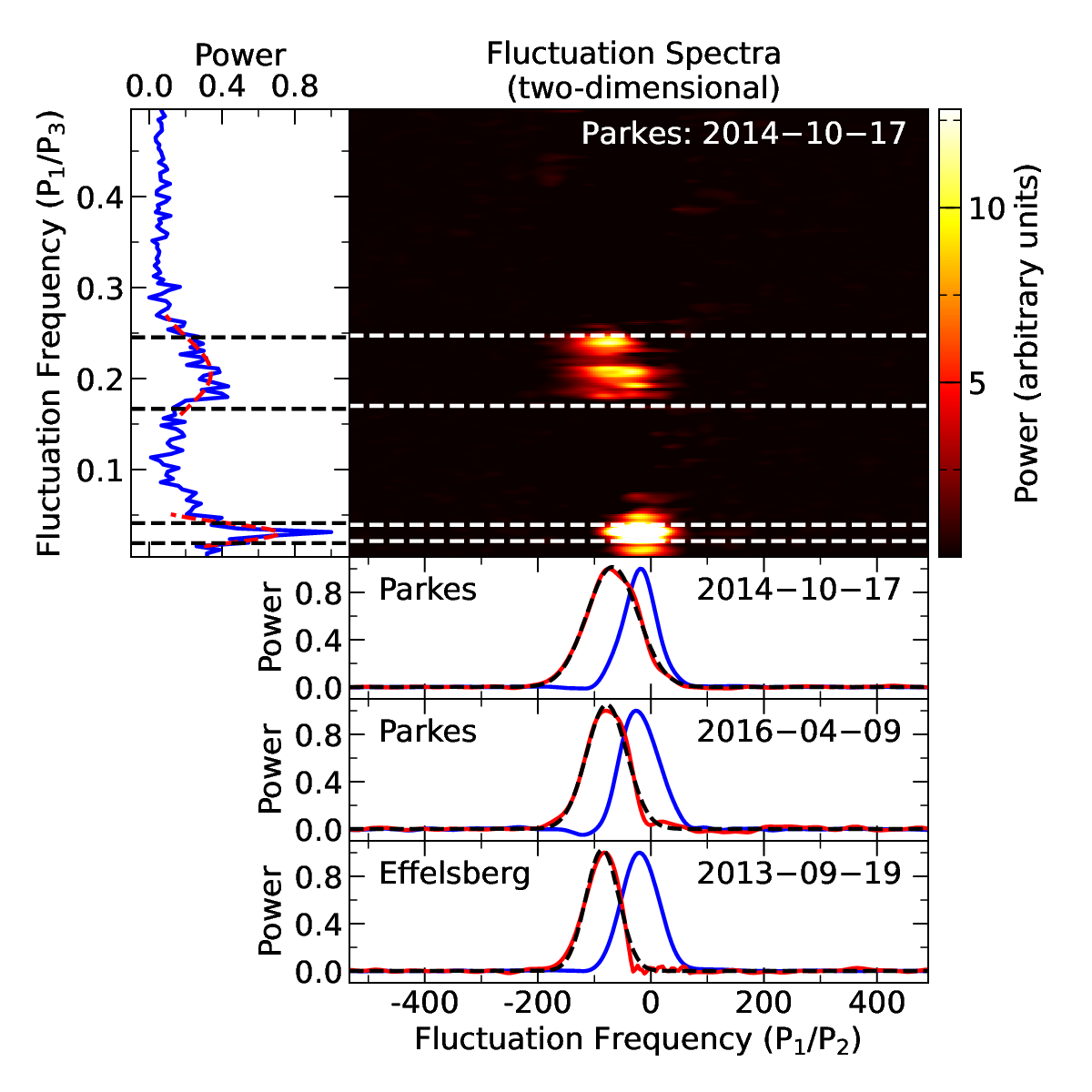}
	\caption{The two-dimensional fluctuation power spectra for PSR
	J1741$-$0840 are presented.
	The main panel displays the full spectrum with a color-coded plot.
	The horizontal integration of the 2DFS yields the left power spectrum, while
	vertical integration between the dashed lines produces the bottom of the
	spectra for three different epochs.
	The black dashed lines in the bottom panels represent the Gaussian fits to
	the modulation power, which is attributed to the the periodicity in the
	subpulse drifting.}
	\label{pic:twodfs}
\end{figure}

In order to further determine whether the periodic modulation originated from
amplitude or longitude variation, the 2DFS is calculated as shown in the main
panel of Figure~\ref{pic:twodfs}.
The 2DFS is integrated both vertically (between dashed lines) and horizontally,
producing the side and bottom panels.
The horizontal axis of the 2DFS represents the repetition frequency of the
modulation pattern along the pulse longitude, expressed as $P_1/P_2$.
This analysis corroborates and clarifies the modulation features 
identified by LRFS and HRFS.
Two modulation features are present: a low-frequency modulation related to the
quasiperiodic nulling, and a high-frequency modulation associated with the
subpulse drifting.
The spectral power of low-frequency modulation feature is perfectly symmetric
about the vertical axis with no biased offset from $P_2$.
This suggests that the subpulses in successive pulses do not shift to later or
earlier pulse longitudes, which is consistent with the statement that the
low-frequency modulation feature is arised from the quasiperiodic nulling.
In contrast, the high-frequency modulation feature reveals a horizontal
structure in the 2DFS, with the spectral power concentrated at negative
frequency values of $P_2$.
This asymmetry of the 2DFS about the vertical axis indicates the appearance of
subpulse drifting and the subpulses tend to drift to earlier pulse longitudes in
successive pulses.
The values of $P_2$ and drift rate, along with their uncertainties are quoted in
Table~\ref{tab:drifting_param}, estimated by fitting a Gaussian to the
spectral power of these features.

\begin{table}
	\centering
	\caption{The details of measurement of subpulse drifting features from PSR J1741$-$0840.}
	\begin{tabular}{ccccc}
		\hline\hline
		Date & $P_3$ & $P_2$ & $D$ \\
		(yyyy$-$mm$-$ss) & (periods) & (deg) & (deg/periods) \\
		\hline
		2013$-$09$-$19 & $5.20^{+0.56}_{-0.46}$ & $4.19^{+0.70}_{-0.53}$ & $0.81^{+0.05}_{-0.03}$ \\
		2014$-$10$-$17 & $4.86^{+0.43}_{-0.36}$ & $5.29^{+2.09}_{-1.17}$ & $1.09^{+0.33}_{-0.16}$ \\
		2016$-$04$-$09 & $4.66^{+0.65}_{-0.51}$ & $4.61^{+1.12}_{-0.75}$ & $0.99^{+0.10}_{-0.05}$ \\
		\hline
	\end{tabular}
	\label{tab:drifting_param}
\end{table}

\begin{figure}
	\centering
	\includegraphics[width=8.0cm,angle=0,scale=1]{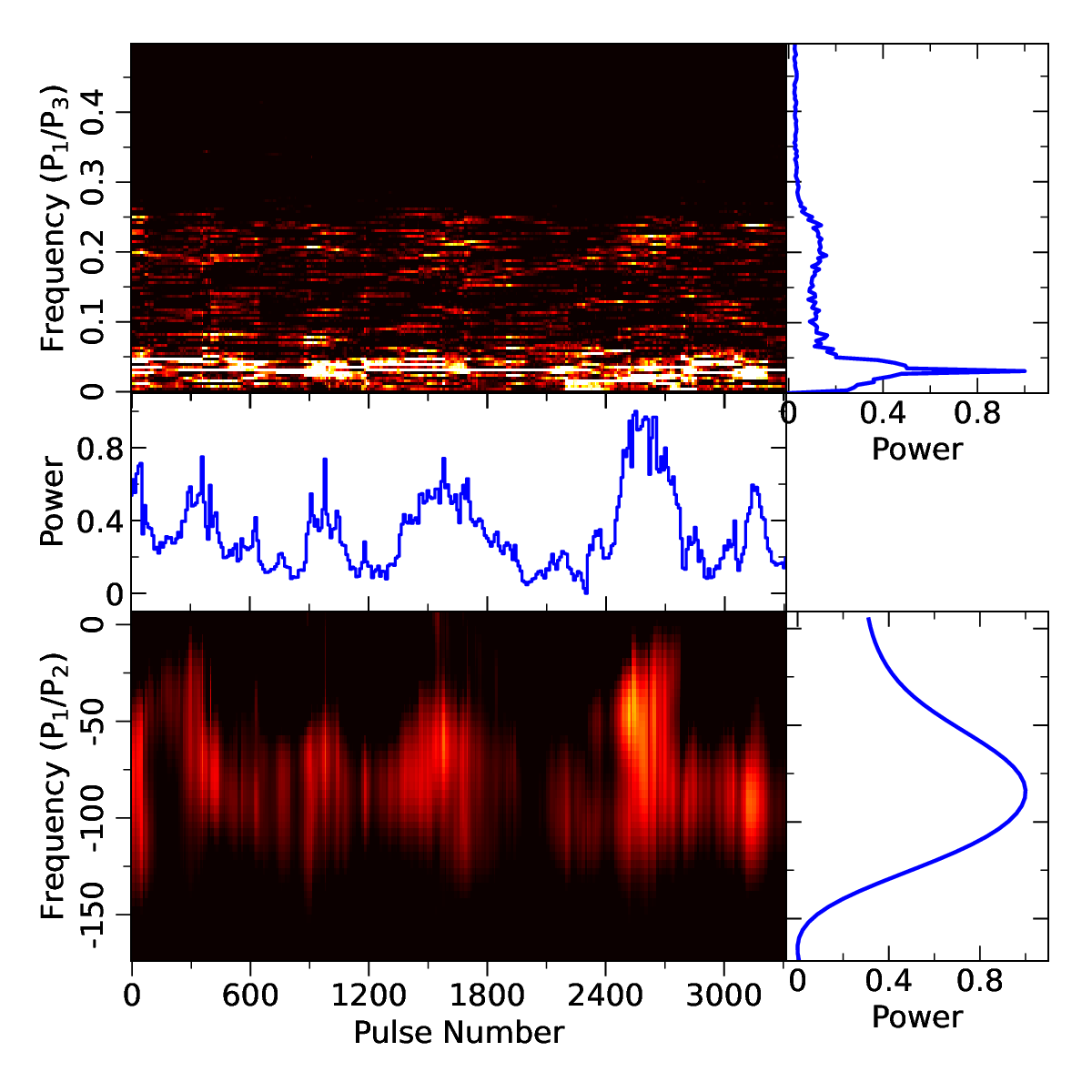}
	\caption{The sliding 2DFS showing the time-dependent spectra by collapsing
	the 2DFS over the $P_2$ and $P_3$ axes for the observations carried out on
	2013 September 19.
	The temporal evolution of the modulation power by integrating the drifting
	feature is shown in the central panel.}
	\label{pic:temevo}
\end{figure}

The modulation features in the fluctuation spectra caused by the subpulse
drifting are broad, which may imply that the periodicity does not appear to be
particularly stable.
In order to explore the stability of the periodicity due to subpulse drifting,
the sliding 2DFS (S2DFS) is used to analyze the time dependence of the values of
$P_3$ and $P_2$.
The upper left panel in Figure~\ref{pic:temevo} shows the time-dependent spectra
by collapsing the 2DFS over the $P_2$ axis for the observations on 2013 September
19.
It is noted that the S2DFS displays a continuously changing frequency of the
modulation pattern within the range of 0.15 to 0.25 cpp, leading to significant
uncertainties in the modulation periodicity.
The lower left panel presents the time-dependent spectra by collapsing the 2DFS
over the $P_3$ axis, which clearly shows the changes in periodic modulation with
time.
The temporal variation of the modulation power by integrating the drifting
feature is shown in the central panel, which presents significant temporal
evolution of the periodic subpulse drifting.
The timescale of the time variability of the modulation pattern is around 500
pulses.

\section{Discussion}
\label{sec:dis}

\subsection{The viewing geometry}
The precise location of the radio emission region within the pulsar
magnetosphere remains a significant uncertainty in the comprehension of pulsar
emission physics.
Generally, the radio emission from pulsars is suggested to be arised from close
to the surface at a height and is confined within the open magnetic field lines.
Calculating the emission altitude through a geometrical approach necessitates
assuming a dipole field configuration, where the viewing geometry must be
specified \citep{Phillips+1992}.
However, the inclination and impact angles cannot be conclusively determined
from RVM fitting, as illustrated in Figure~\ref{pic:pol_prof_corner}.
An alternative method with notable advantages for estimating emission heights
was initially discussed by \citet{Blaskiewicz+etal+1991}.
This method posits that the inflection point exhibits a time lag ($\Delta\phi$)
relative to the midpoint of the profile caused by the relativistic aberration
and retardation effects due to the rapid rotation of the magnetosphere.
The emission height at the observing frequency is given by
\begin{equation}
	h_{\rm{em}} = \frac{P_1 c \Delta\phi}{8\pi},
\end{equation}
where $c$ is the speed of light.
For PSR J1741$-$0840, the profile centroid is identified as the position of the
central component.
Thus, the emission height is derived to be 1021$\pm$55 km.
The half-opening angle of the emission beam is as well given \citep{Rankin+1990} by 
\begin{equation}
	\rho = 3 \sqrt{\frac{\pi h_{\rm{em}}}{2 P_1 c}},
\end{equation}
which yields $\rho \sim 9^\circ$.
In turn, coupled with the observed pulse width, the half-opening angle can be
expressed via the emission geometry \citep{Gil+etal+1984} as
\begin{equation}
	\cos\rho = \cos\alpha\cos\zeta+\sin\alpha\sin\zeta\cos(W_{10}/2),
\end{equation}
where $W_{10}$ is the pulse width measured at 10\% of the peak flux.
In the $\alpha-\beta$ plane as shown in Figure~\ref{pic:pol_prof_corner}, the
white contours of constant emission height make an almost orthogonal cut through
the `banana', and this contour intersects the reduced $\chi^2$ of the RVM
fitting at $\alpha=81.09^\circ$ and $\beta=3.26^\circ$, corresponding to an
outer line of sight.

It is noted that pulsar J1741$-$0840 presents trailing partial cone, where only
the trailing part of the cone is illuminated occasionally.
\citet{Johnston+etal+2023} found that the trailing partial cones constitute only
a tiny fraction of the entire pulsar population.
Additionally, the emission heights for the majority of the radio pulsar
population are less than 1000 km, and clear evidence for an evolution of the
magnetic axis towards the spin axis wit time is proposed.
PSR J1741$-$0840 is a relatively old pulsar with a spin-down age of 14.2 Myr.
The determined large value of $\alpha$ may suggest that the axis of the dipolar
magnetic field is moving towards the stellar equator, evidence for which is also
apparent in the Crab pulsar \citep{Lyne+etal+2013}.

\subsection{Implication of quasiperiodic nulling and subpulse drifting}
PSR J1741$-$0840 shows the presence of periodic modulation features associated
with nulling as well as subpulse drifting.
The phenomenon of subpulse drifting is generally explained in the seminal polar
gap theory originally proposed by \citep{Ruderman+Sutherland+1975}.
Each subpulse is associated with a subbeam in the pulsar magnetosphere.
Then the drift in the subpulse is attributed to the rotation of these subbeams
around the pulsar's magnetic axis.
By assuming the subbeams are equally spaced, the angular separation ($\eta$)
between two adjacent subbeams along the magnetic axis is expressed as
\begin{equation}
	\sin(\eta/2) = \frac{\sin(P_2/2)\times\sin(\alpha+\beta)}{\sin(\beta)}.
\end{equation}
The magnetic azimuth angles are calculated to be 91.63$^\circ$.
The corresponding numbers of subbeams are estimated to be around 4.
The circulation time that a subbeam takes for one rotation around the magnetic
axis is derived to be 19.29 periods.

The average pulse profile is generally generated by integrating some
hundreds of individual pulses, so a rich diversity of behaviour among the
individual pulses is concealed in the average profile.
An individual pulse often comprises multiple components characterized by
subpulses, which are considered basic components of the pulse profile.
An individual subpulse is suggested to be the radiation from an isolated
location within the emission region covered by the integrated profile.
Therefore, the integrated pulse profile represents the statistical distribution of 
subpulses across a range of longitude, combined with their characteristic widths and the
probability distribution of their intensities which determine the shape of the
integrated profile.
The integrated pulse profile of PSR J1741$-$0840 comprises of four emission
components, which corresponds to the number of subbeams derived from subpulse
drifting phenomenon.

The nature of nulling and its quasiperiodic behaviour can be explained within
the context of the rotating carousel model as well.
The periodic nulls are proposed to be resulted from the precession of the
subbeams as the missing or empty line-of-sight traverse between the subbeam
system \citep{Basu+etal+2017}.
Thus, the short-duration nulls are also termed as pseudo nulls
\citep{Herfindal+Rankin+2009}.
However, this explanation is no longer consistent for pulsars with long-duration
nulls \citep{Basu+etal+2017}.
Although PSR J1741$-$0840 presents clear evidence of quasiperiodic nulling, the
lack of drifting phase memory across nulls and the none-detection of partial
nulls suggest that the mission beamlets in the rotating carousel model is
implausible to generate the qusiperiodic nulling shown in PSR J1741$-$0840
\citep{Gajjar+etal+2017}.
Therefore, \citet{Grover+etal+2024} proposed that pulse nulling occurs independent of
subpulse drifting, and its quasiperiodic nature is attributed to other factors.

\section{Conclusions}
\label{sec:con}
We have conducted a detailed analysis of new observations of PSR J1741$-$0840
with the Parkes and Effelsberg radio telescopes at the $L$ band.
The pulsar spends roughly 23\% of the total observation in the nulling state.
The durations of null and burst modes can be well described by power-law
distributions.
Using fluctuation spectral analysis, PSR J1741$-$0840 shows the presence of
periodic modulations at two different periodicities, 32$P_1$ corresponding to
the quasiperiodic nulling, and 5$P_1$ arising due to the phenomenon of subpulse
drifting.
The following multi-frequency observations are required to characterize the spectral
development of the pulsar's profile, investigate the frequency dependence of
subpulse drifting and quasiperiodic nulling, determine the frequency evolution of
opening angle, and further make inferences about the emission geometry.

\section*{Acknowledgements}
\addcontentsline{toc}{section}{Acknowledgements}
The authors would like to thank the anonymous referee for useful comments
that helped improve the paper.
This work is partially supported by the National Key Research and Development
Program of China (No. 2022YFC2205203), the Major Science and Technology Program of
Xinjiang Uygur Autonomous Region (No. 2022A03013-1), the National Natural Science
Foundation of China (NSFC grant Nos. 12303053, 12288102, 12041301, U1838109,
12203094), the open project of the Key Laboratory in Xinjiang Uygur Autonomous
Region of China (2023D04058), the Chinese Academy of Sciences Foundation of the
young scholars of western (2020-XBQNXZ-019).
ZGW is supported by the 2021 project Xinjiang Uygur autonomous region of China
for Tianshan elites, the Youth Innovation Promotion Association of CAS under
No. 2023069, and the Tianshan Talent Training Program (No. 2023TSYCCX0100).
YHX is supported by the CAS ``Light of West China" Program.
HGW is supported by the 2018 project of Xinjiang Uygur autonomous region of
China for flexibly fetching in upscale talents.
WH is supported by the CAS Light of West China Program No. 2019-XBQNXZ-B-019.
This work makes use of observations conducted with the 100 m radio telescope in
Effelsberg owned and operated by the Max-Planck Institut f$\ddot{\rm{u}}$r Radioastronomy.
The Parkes radio telescope, `Murriyang', is part of the Australia Telescope
National Facility, which is funded by the Australian Government for operation as
a National Facility managed by CSIRO.
We thank the observers in setting up the observations.

\bibliographystyle{aasjournal}
\bibliography{bibtex}

\end{document}